\documentclass{llncs} 

\usepackage{xspace}
\usepackage{graphicx}
\usepackage{subfigure}
\usepackage{url}
\usepackage{todonotes}
\usepackage[capitalise]{cleveref}

\newcommand{\sys}{VAIM\xspace}

\title{\sys: Visual Analytics \\for Influence Maximization\thanks{Research of WD, GL and FM partially supported by: $(i)$ MIUR, grant 20174LF3T8 ``AHeAD: efficient Algorithms for HArnessing networked Data'', $(ii)$ Dip. di Ingegneria - Universit\`a degli Studi di Perugia, grant RICBA19FM: ``Modelli, algoritmi e sistemi per la visualizzazione di grafi e reti''. Research of AA and SM partially supported by TU Wien ``Smart CT'' research cluster.}}

\author{Alessio Arleo\inst{1}\orcidID{0000-0003-2008-3651} \and
		Walter Didimo\inst{2}\orcidID{0000-0002-4379-6059} \and
        Giuseppe Liotta\inst{2}\orcidID{0000-0002-2886-9694} \and
        Silvia Miksch\inst{1}\orcidID{0000-0003-4427-5703} \and
        Fabrizio Montecchiani\inst{2}\orcidID{0000-0002-0543-8912}
}

\authorrunning{A. Arleo, W. Didimo, G. Liotta, S. Miksch, F. Montecchiani}

\institute{
	TU Wien, Austria\\
	\email{name.surname@tuwien.ac.at}
	\and
	Universit\`a degli Studi di Perugia, Italy\\
	\email{name.surname@unipg.it}
}

\begin{document}

\maketitle

\begin{abstract}
In social networks, individuals' decisions are strongly influenced by recommendations from their friends and acquaintances. The \emph{influence maximization} (IM) problem asks to select a \emph{seed set} of users that maximizes the influence spread, i.e., the expected number of users influenced through a stochastic diffusion process triggered by the seeds. In this paper, we present \sys, a visual analytics system that supports users in analyzing the information diffusion process determined by different IM algorithms.
By using \sys one can: (i) simulate the information spread for a given seed set on a large network, (ii) analyze and compare the effectiveness of different seed sets, and (iii) modify the seed sets to improve the corresponding influence spread.

\end{abstract}

\section{Introduction}\label{se:introduction}

People in social networks influence each other in both direct and indirect ways, through a mechanism often known as the \emph{word-of-mouth effect} (see, e.g.,~\cite{DBLP:conf/kdd/KempeKT03,DBLP:journals/toc/KempeKT15}).
For this reason social networks are becoming the favorite venue where companies advertise their products/services and where politicians run their campaigns. The \emph{influence maximization} (IM) problem asks to select a \emph{seed set} of users that maximizes the influence spread, i.e., the expected number of users positively influenced by an information diffusion process triggered by the seeds and that spreads through the network according to some stochastic model.
We refer the reader to the works by Guille et al.~\cite{DBLP:journals/sigmod/GuilleHFZ13} and by Li et al.~\cite{DBLP:journals/tkde/LiFWT18} for surveys about influence maximization and information diffusion in social networks.

\begin{figure}[tb]
	\centering
	\includegraphics[width=1\columnwidth]{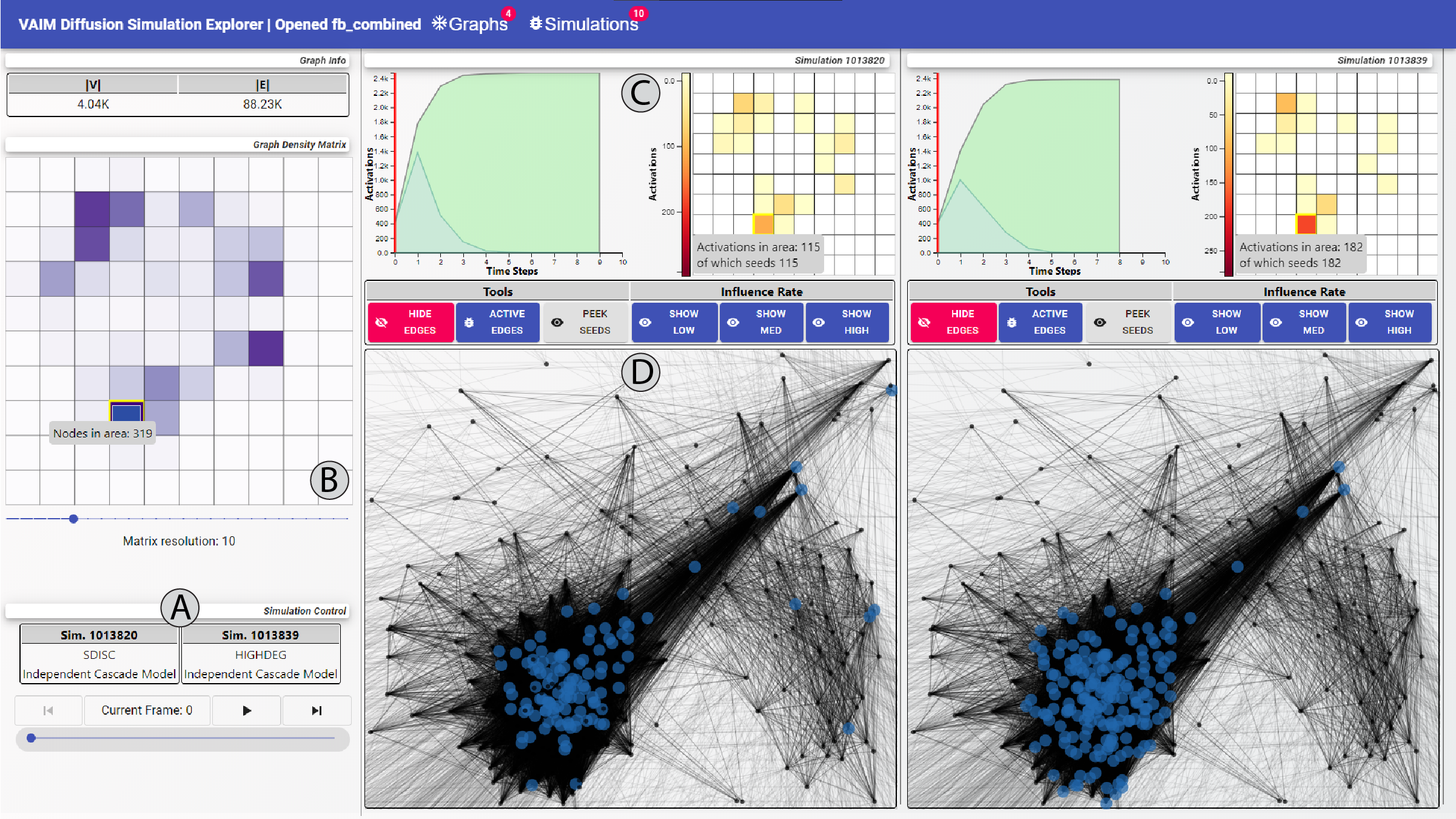}
	\caption{\sys's visual interface,
	at $t=0$ of the case study~in \cref{se:eval}. Its components are marked as follows: A) Simulation control, B) Density Matrix view,  C) Diffusion Matrix view, D) Node-link view.}\label{fi:cs1-1}
\end{figure}

Analyzing and engineering an IM algorithm is a demanding task; as reported by Arora et al.~\cite{DBLP:conf/sigmod/AroraGR17}, there is no single state-of-the-art technique for IM.
Under the most common diffusion models, finding the optimal seed set in a network is known to be an NP-hard problem~\cite{DBLP:conf/kdd/KempeKT03}.
Besides the problem hardness, being the information diffusion process stochastic, even the evaluation of influence spread of any seed set is computationally complex~\cite{DBLP:conf/kdd/ChenWW10}, which makes the design of scalable and effective IM algorithms a great challenge that motivated a large and still increasing body of literature~\cite{DBLP:journals/tkde/LiFWT18}.
%
%
In this context, we want to exploit the power of information visualization to support expert users in analyzing, evaluating, and comparing IM algorithms. Our main contributions are as follows.

\smallskip\noindent $(i)$ We present \sys, a system that provides facilities to simulate an information diffusion process over a given network and problem-oriented visual analytics (VA) tools to explore the related data (\cref{se:design}).
\sys has a modular architecture that currently includes some of the most popular IM algorithms and information diffusion models.
An interface with multiple coordinated views makes it possible to visually compare and analyze the performance of a diffusion model over potentially very large networks and for different choices of the seed sets (i.e., for different IM algorithms). The user can interactively modify the seed set and iterate the process until a satisfying spread is achieved.

\smallskip\noindent $(ii)$ The effectiveness of \sys is evaluated through a case study (\cref{se:eval}). We show how tacking advantage of \sys for (a) comparing different seed selection algorithms on the same network, and (b) improving the seed selection by either a manual or a system-assisted modification of the initial seed set.


\medskip\noindent{\bf Related work.}
There are several visualization systems designed to analyze information diffusion processes in social networks.
TwitInfo~\cite{DBLP:conf/chi/MarcusBBKMM11,DBLP:journals/sigmod/MarcusBBKMM11} aggregates tweets in the spatial, temporal, and event dimensions supporting the exploration of event propagation processes. Whisper~\cite{DBLP:journals/tvcg/CaoLSLLQ12} exploits a flower-like visualization for real-time monitoring of the diffusion of a given topic, highlighting the spatio-temporal information of the process over the world. OpinionFlow~\cite{DBLP:journals/tvcg/WuLYLW14} uses Sankey graphs and density maps to visually summarize opinion diffusion processes. FluxFlow~\cite{DBLP:journals/tvcg/ZhaoCWSLC14} adopts a timeline visualization to analyze anomalous information diffusion spreading. D-Map~\cite{DBLP:conf/ieeevast/ChenCWLYCW16} collects data from Sina Weibo and offers a map-based ego-centric visualization to reveal dynamic patterns of how people are involved and influenced in a diffusion process. SocialWave~\cite{DBLP:journals/tist/SunTPLW18} uses abstract visualizations to explore and analyze spatio-temporal diffusion of information.
More approaches are elaborated in Chen et al.~\cite{DBLP:journals/cgf/ChenLY17}.
All these approaches are designed to reveal different facets of information diffusion processes and they often rely on geographical and other user-related information.
On the other hand, they neither support the user in analyzing the impact of the seeds (which in fact may be unknown) and of the network structure in terms of influence spread, nor offer simulation tools to experiment different diffusion models.


Long and Wong~\cite{DBLP:conf/icdm/LongW14} introduce Visual-VM, a visualization system for viral marketing. Similar to \sys, Visual-VM allows users to simulate stochastic diffusion processes and to visually analyze their output. However, Visual-VM offers a simple visual interface, which strongly relies on geographical information to lay out the network. The networks analyzed with \sys may come from diverse scenarios and may not contain geographical information about users.

Finally, Vallet et al.~\cite{DBLP:journals/corr/ValletKPM15,DBLP:conf/ieeevast/ValletPM14} present a visualization framework to compare different diffusion models based on a common set of graph rewriting rules. Different from \sys, the work of Vallet et al. does not focus on comparing different IM algorithms and it is mainly tailored to networks of small or medium size.

\medskip\noindent{\bf Background and notation.} We model a social network as a directed graph $G=(V,E)$. 
A \emph{diffusion model} $M$ captures the stochastic diffusion process among the vertices of $G$. During the process, a vertex $v \in V$ can be either \emph{active} or \emph{inactive}.  The \emph{influence spread} of a seed set $S$, denoted by $\sigma_{G,M}(S)$, is the expected number of active vertices  once the diffusion process (over the graph $G$ and under the model $M$) terminates. More formally, the IM problem asks for a set $S^* \subseteq V$ of at most $0<k \le |V|$ seeds that maximizes the influence spread, i.e., $S^* = arg\,max\{\sigma_{G,M}(S) | S \subseteq V \wedge |S|\le k\}.$
%
%
One of the most commonly used diffusion models is the \emph{Independent Cascade} (IC)~\cite{DBLP:journals/tkde/LiFWT18}. Other models (such as the Linear Threshold model) make use of additional parameters but do not differ significantly in terms of the underlying iterative framework. In the IC model, a diffusion instance unfolds through an iterative process: In step $0$, only the seed vertices are active; in step $j>0$, each vertex $u$ activated at step $j-1$ will activate each of its inactive neighbors $v$ with probability $0 \le p(u,v) \le 1$. The process halts when no more vertices can be activated.
Unfortunately, the IM problem is \textsc{NP}-hard under the IC model, as well as under other models~\cite{DBLP:conf/kdd/KempeKT03}. 
For a broader discussion refer to~\cite{DBLP:journals/sigmod/GuilleHFZ13,DBLP:journals/tkde/LiFWT18}.

\section{\sys Design}\label{se:design}

The design of \sys relies on the ``Data-Users-Tasks'' model proposed in~\cite{miksch2014matter}.

\medskip\noindent\textbf{Data.}
To estimate the influence spread of a seed set, we rely on a simulation-based approach. To obtain statistically relevant data, the simulation is repeated multiple times. Each single repetition is a time-dependent process taking as input a graph and a set of seeds. Hence, the data model of \sys includes the input network, and set-typed temporal data represented by the active set of vertices and edges at every timestamp of the simulated diffusion process.

\medskip\noindent\textbf{Users.}
\sys targets a single class of expert users. Those users are knowledgeable in their own application domain and in the use of visual analytics tools. Also, they are interested not only in the resulting influence spread, but also on how the structure of the network influences the diffusion process.
%


\medskip\noindent\textbf{Tasks.}
\sys is designed to support the following user tasks:

\medskip\noindent  \textsf{T1 Simulate.} It should be possible to simulate a diffusion process on a given network, with the seeds from an IM algorithm, under a given diffusion model.

\medskip\noindent  \textsf{T2 Evaluate.} The user should be allowed to visually analyze both the quality of spread of a seed set and the impact of the network structure on the diffusion process, such as areas with a higher rate of active nodes, isolated areas, etc. The user can fast forward, rewind, and pause the process animation. 

\medskip\noindent  \textsf{T3 Compare.} It should be possible to visually compare the performance of different seed sets computed by different IM algorithms.

\medskip\noindent  \textsf{T4 Feedback.} The user should be facilitated in modifying the seed set and iterate the simulate-evaluate-compare process.

\subsection{Visualization design}\label{sse:visualization}
The visualization design adopts an overview+detail approach. The interface is organized as a dashboard with multiple coordinated views (see also \cref{fi:cs1-1}). The chosen colour schemes and palettes are colorblind friendly~\cite{harrower2003colorbrewer}.

\medskip\noindent -- \textsf{Simulation control (Fig. 1-A).}
Here the user can set different parameters about the diffusion process, such as
the stochastic model and the number of iterations (Task~\textsf{T1}). 

\medskip\noindent -- \textsf{Density matrix view (Fig. 1-B)}. The main purpose of this view is to provide an overview of the network structure in a scalable manner. This is achieved with a  simplified matrix visualization, which is obtained by firstly computing a node-link layout of the whole (potentially very large) network with some fast algorithm, such as centralized or distributed force-directed techniques (e.g.,~\cite{DBLP:journals/isci/ArleoDLM17,DBLP:journals/tpds/ArleoDLM19,DBLP:reference/crc/Kobourov13}), and then by slicing the plane into cells. The color intensity of each cell reflects the number of nodes inside. The size of the matrix can be increased or decreased through a simple slider. Hovering with the mouse on a cell, the number of nodes in that cell is reported. 

\medskip\noindent -- \textsf{Diffusion matrix view (Fig. 1-C)}. It allows users to visually compare multiple simulations  over the same network. A legend below it shows the considered IM algorithms. Each simulation is conveyed using a distinct matrix visualization whose cells' colors vary in a \textit{YlOrRd} scale (yellow to orange to red) and reflect the number of active nodes in the corresponding area of the network. Notably, the density and diffusion matrices have the same set of cells, so to facilitate comparisons and associations among them. Similarly as for the density matrix view, the computation and the rendering of this view must be fast enough to allow the visualization of multiple simulations over large networks.
At the left side of each diffusion matrix, the \textsf{process trend chart} is a plot with two curves showing, for each iteration, the number of new nodes activated in that iteration and the cumulative number of nodes activated up to that iteration. A red vertical segment indicates the currently selected iteration. \sys can animate the diffusion process over time. Other facilities allow users to highlight those cells containing some seeds, or whose influence rate is low ($<30\%$), medium ($[30\%, 60\%]$), or high ($> 60\%$). Clicking on a cell,  its influence rate is shown and a list of nodes that can be either removed or promoted as seeds is suggested, based on node degrees and influence rate.

\medskip\noindent -- \textsf{Node-link view (Fig. 1-D).} Below each diffusion matrix, there is a panel in which a detailed node-link diagram of a portion of the network can be visualized. This portion can be freely chosen by the user through a brushing selection of any group of $k \times h$ cells in the density matrix. The combination of the node-link view with the two matrix views described above is particularly useful for very large networks, for which detailed visualizations are feasible only for small portions. In the diagram, blue nodes represent seeds while dark red nodes and edges represent the active elements at the considered time instant (\cref{fi:cs1-2}). The user can hide all edges or leave only the active ones.

\smallskip The three views together are designed to support Tasks~\textsf{T2},~\textsf{T3}~and~\textsf{T4}.


\section{Evaluation and Discussion}\label{se:eval}

\begin{figure}[p]
	\centering
	\subfigure[]{\includegraphics[width=1\columnwidth]{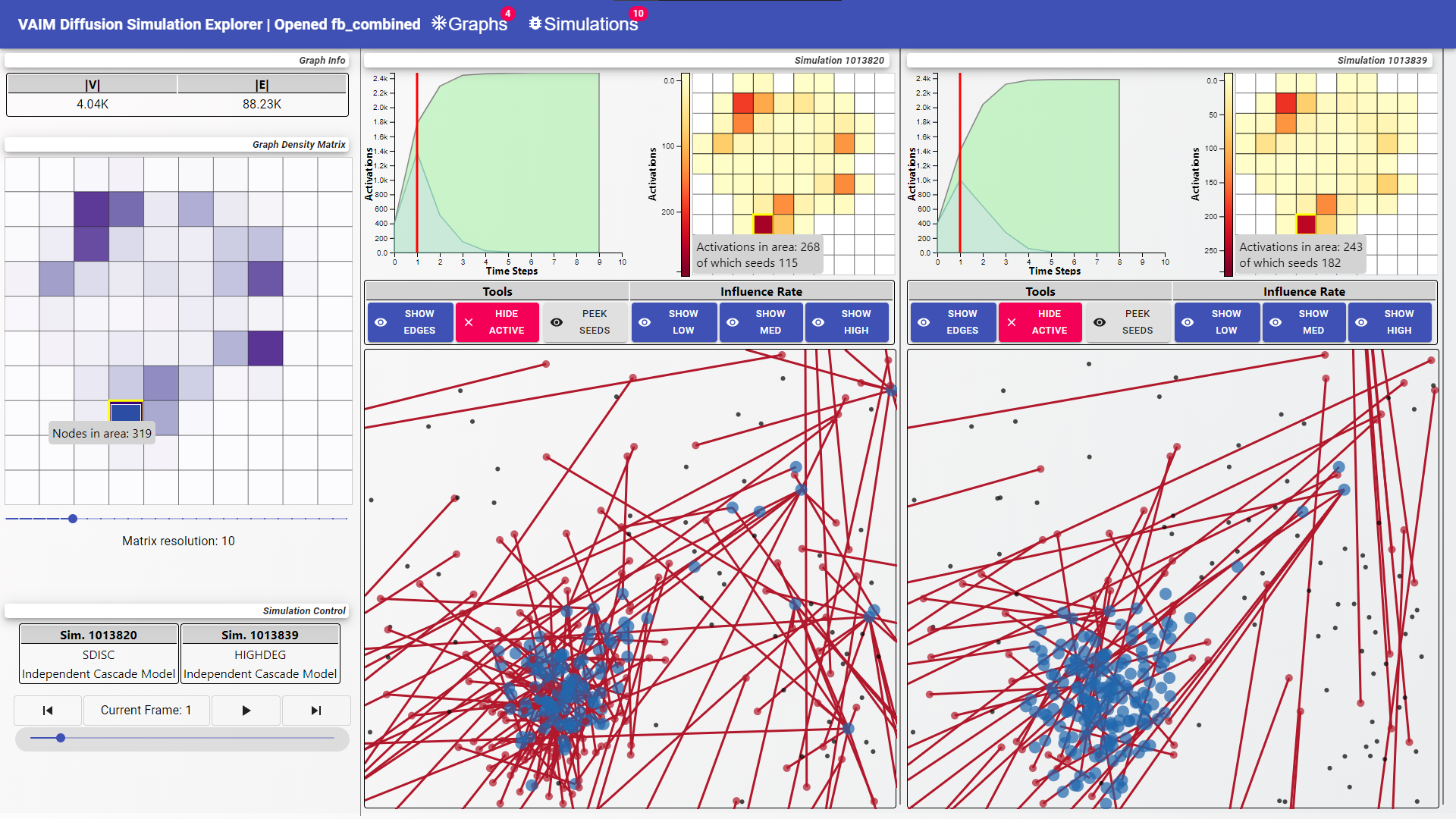}\label{fi:cs1-2}}
	\subfigure[]{\includegraphics[width=1\columnwidth]{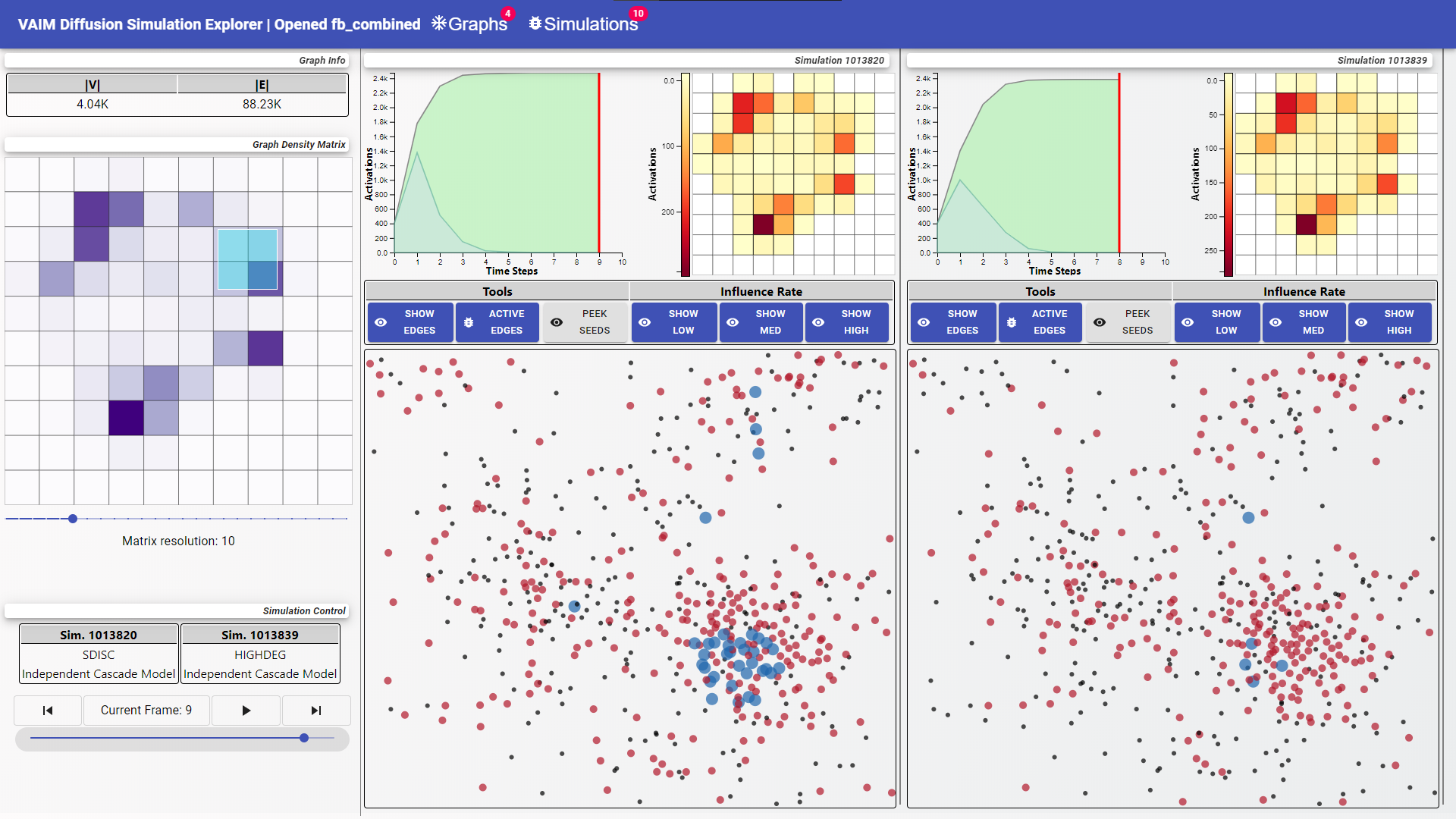}\label{fi:cs1-3}}
	\caption{Snapshot of \sys after (a) the first iteration of the diffusion process, and (b) at the end of the diffusion process.}
\end{figure}

We discuss an evaluation of \sys based on the following case study (see the appendix for an additional case study). The input is the \textsf{fb-combined} social network, extracted from Facebook~\cite{leskovec2012learning}, having $4,039$ nodes and $88,234$ edges ({\small \url{https://snap.stanford.edu/data/}}). We simulated an IC diffusion process (\textsf{T1}), using two seed sets of $400$ nodes each, computed by two popular IM algorithms: \textsf{HIGHDEG}~\cite{DBLP:conf/kdd/KempeKT03,DBLP:journals/toc/KempeKT15} and \textsf{SDISC}~\cite{10.1145/1557019.1557047}, based on degree centrality and discount, respectively. We compared and evaluated (\textsf{T2} and~\textsf{T3}) the performance of the two diffusion processes. \cref{fi:cs1-1} shows a snapshot of the interface at the beginning of each diffusion process. The process trend charts reveal that \textsf{SDISC} leads to a higher number of active nodes in fewer iterations. By exploring the diffusion matrices we can observe a different distribution of the seeds selected by the two IM algorithms. For example, focusing on the densest cell of the network (which can be easily spotted in the density matrix), we see that \textsf{HIGHDEG} (the right-side simulation) concentrates a higher number of seeds than \textsf{SDISC} (the left-side simulation) in that cell (182 seeds of \textsf{HIGHDEG} vs 115 seeds of \textsf{SDISC}), while putting relatively fewer seeds in sparser cells. Also, within the densest cell, \textsf{SDISC} distributes the seeds more uniformly than \textsf{HIGHDEG}.
\cref{fi:cs1-2} shows the processes at the next iteration, and still focuses on the densest cell. Despite the smaller number of seeds, \textsf{SDISC} yields a higher number of newly active nodes (red nodes) in that cell (268 of \textsf{SDISC} vs 243 of \textsf{HIGHDEG}). Also, the greater number of red edges (those used by the diffusion process) exiting the cell, reveals a higher influence of the nodes of this cell towards nodes outside it.
\cref{fi:cs1-3} shows the end of the processes. Using the influence rate function, we observe that the cells selected from the density matrix have a smaller number of active nodes with \textsf{HIGHDEG} than with \textsf{SDISC}. Looking at the node-link view for these cells (edges are hidden), this seems to be caused by the very small number of seeds that \textsf{HIGHDEG} placed in this portion of the network. The above discussion helps understanding how the seeding strategy adopted by \textsf{SDISC} leads to better performance, which corroborates the results of an experimental analysis performed on a collaboration graph presented in~\cite{10.1145/1557019.1557047}.

In order to improve the information spread of \textsf{SDISC} (\textsf{T4}), \sys suggested 20 nodes (with smallest degree in the cell with highest influence rate) to be removed from the original seed set and  20 nodes (with highest degree in the cell with lowest influence rate) to be promoted as seeds. We modified the seed set accordingly and we simulated again the diffusion process. The new process lead to~2\%~more~of~active~nodes.

\section{Conclusion and Future Work}\label{se:conclusions}

We discussed the use of visual analytics to support the analysis and fine tuning of IM strategies. We plan to extend the system with features such as edge bundling to mitigate edge clutter in the node-link view. We will also implement new diffusion models, together with ad-hoc views to explore the additional parameters of these models. Considering networks with node and edge attributes (e.g., geo-locations) is also an interesting direction. Finally, we want to further evaluate \sys with more case studies and experiments, in particular to test its scalability (both in terms of simulation and visualization) to~very~large~networks.

\clearpage

\bibliography{vaim_arxiv}
\bibliographystyle{splncs04}

\clearpage

\appendix

\section*{Appendix}\label{ap:appendix}

\section{Additional Case Study}

\begin{figure}[h]
	\centering
	\includegraphics[width=1\columnwidth]{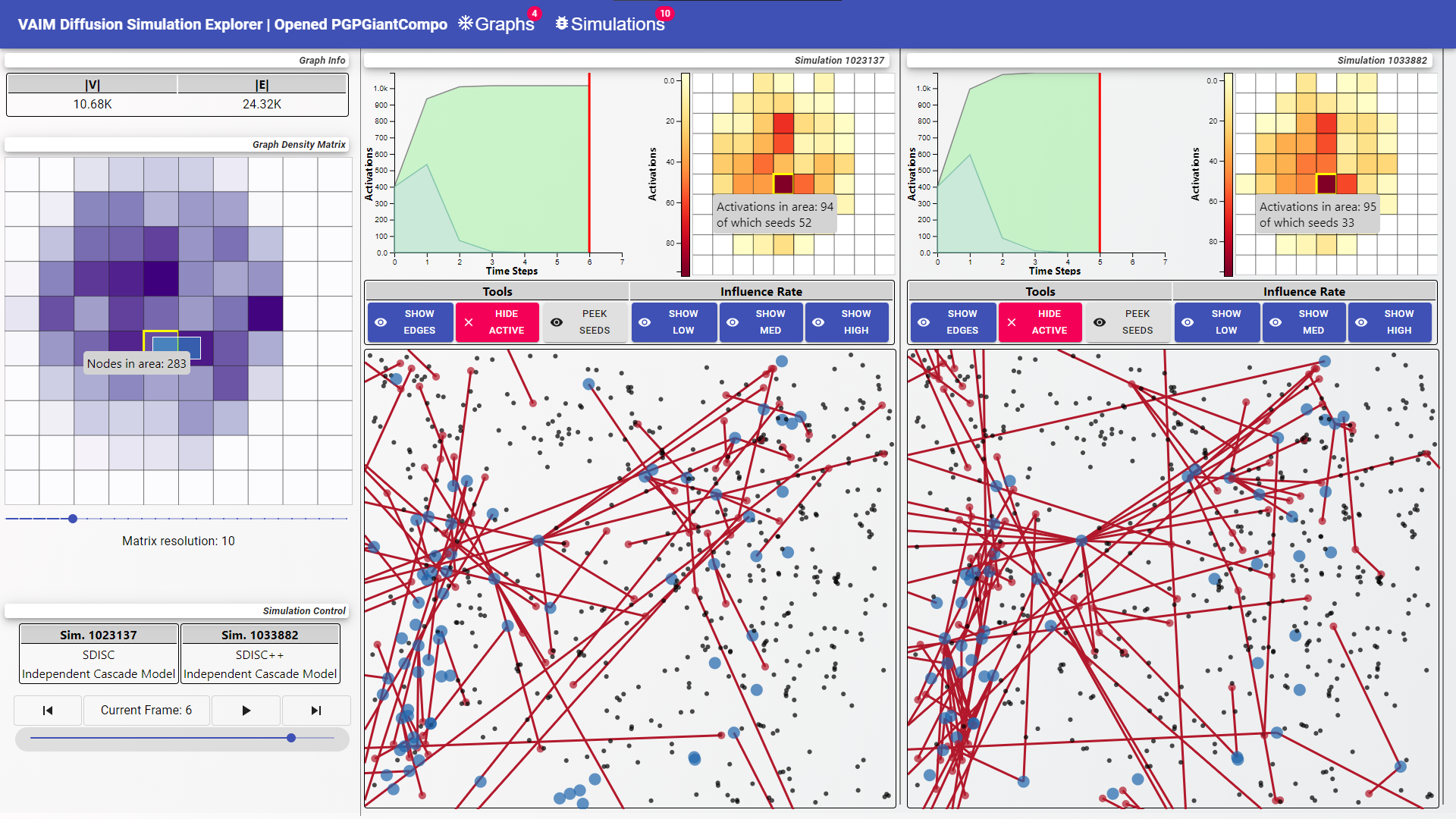}
	\caption{Snapshot of \sys at the end of the diffusion process for the second case study. The system suggested a change in the seed distribution (right side) over the one computed by SDISC (left side): in particular, removing seeds from the highlighted high-rate region did not harm the local spread performance, which increased by two units instead.}\label{fi:cs2-1}
\end{figure}

We performed a second case study on the email-exchange network \textsf{pgp-giant-compo} ({\small\url{http://networkrepository.com/}}), having $n=10,680$ nodes and $m=24,320$ edges. As in the first case study, we simulated an IC diffusion process with a seed set of $400$ nodes computed by the SDISC IM algorithm. 

By using the influence rate function, we identify the cell with the lowest number of activations. The system suggests a list of 20 nodes to add to the seed set, chosen among those with the highest degree located in the selected cell. Afterwards, we ask \sys to provide a list of seeds to remove from the SDISC selection. These are picked among the nodes with lowest degree placed in the cell with the highest activation rate. We ran the simulation again with the modified seed set (that still holds the same number of elements) and we obtain an average increase in the spread performance of 1\% (around 100 nodes).

In Figure \ref{fi:cs2-1} we report the comparison of the two diffusion processes at the end of the simulation. By looking at the two diffusion matrices, we could see how the updated seed set achieves higher activation rates on the upper left side of the network, where the new seeds were placed. Moreover, the seeds removed from the most dense cell (in terms of number of activated nodes) did not harm the spread in that area; on the contrary, we observed a small increase in the number of activations.

\end{document}